\documentclass[%
 reprint,
 amsmath,amssymb,
 aps,
]{revtex4-1}

\usepackage{graphicx}% Include figure files
\usepackage{dcolumn}% Align table columns on decimal point
\usepackage{bm}% bold math
\usepackage{subfigure}

\begin{document}

\preprint{APS/123-QED}

\title{The Impact of Information Dissemination \\on Vaccination in Multiplex Networks}% Force line breaks with \\
%\footnotetext{$^\ast$Corresponding Author. Email: xxx@gmail.com(M.Liu).}

\author{Xiao-Jie Li}%
\author{Cong Li}
\email{cong\_li@fudan.edu.cn}
\author{Xiang Li}

\affiliation{ \mbox{Adaptive Networks and Control Lab, Department of Electronic Engineering, }\\
\mbox{Research Center of Smart Networks and Systems, School of Information Science and Engineering}\\
Fudan University, Shanghai 200433, China.}

\begin{abstract}
The impact of information dissemination on epidemic control is essentially subject to individual behaviors. Unlike information-driven behaviors, vaccination is determined by many cost-related factors, whose correlation with the information dissemination should be better understood. To this end, we propose an evolutionary vaccination game model in multiplex networks by integrating an information-epidemic spreading process into the vaccination dynamics, and explore how information dissemination influences vaccination. The spreading process is described by a two-layer coupled susceptible-alert-infected-susceptible (SAIS) model, where the strength coefficient between two layers is defined to characterize the tendency and intensity of information dissemination. We find that information dissemination can increase the epidemic threshold, however, more information transmission cannot promote vaccination. Specifically, increasing information dissemination even leads to a decline of the vaccination equilibrium and raises the final infection density. Moreover, we study the impact of strength coefficient and individual sensitivity on social cost, and unveil the role of information dissemination in controlling the epidemic with numerical simulations.
\begin{description}
\item[PACS numbers]
 {02.50.Le, 89.65.-s, 89.75.Fb}
\end{description}
\end{abstract}

\pacs{Valid PACS appear here}% PACS, the Physics and Astronomy
\maketitle

%\tableofcontents

\section{\label{Sec_1}Introduction}

The epidemic control has been studied many decades  \cite{pastor2001epidemic,theodorakopoulos2013selfish,chen2017spread,pastor2015epidemic}, since the outbreak and propagation of virus may cause tremendous damage and bring huge (economic) losses. Various models, such as the susceptible-infected-susceptible (SIS) model \cite{zhang2014susceptible,li2012susceptible} and the susceptible-infected-recovered (SIR) model \cite{wang2016identifying}, have been used to describe epidemic spreading processes. The understanding of disease-behavior dynamics motivates more and more efforts to explore the epidemic dynamics beyond such models \cite{wang2015coupled,2017minimizing,Vaccinating2017}. Individual behaviors, such as wearing masks and washing hands, which may reduce the susceptibility to infection, can be triggered by the awareness (information) diffusion. Funk $\emph{et}$ $\emph{al.}$ \cite{funk2009spread} studied how awareness impacts the virus propagation in a well-mixed population, finding that the awareness diffusion can reduce the outbreak range, but cannot affect the epidemic threshold. Similarly, Wu $\emph{et}$ $\emph{al.}$ \cite{wu2012impact} explored the impact of three different kinds of awareness on the epidemic spread in a scale-free networked population. However, single-layer networks provide a limited representation of complex systems \cite{de2016physics}. The efforts in \cite{funk2009spread,wu2012impact} may fail to involve the realistic scenario where the information and virus spread via different networks simultaneously.

Recently, multiplex networks representing social interactions at different contexts, e.g. individuals transmit information through an online social network, and at the same time an epidemic propagates among the individuals on a  physical contact network, have been studied in \cite{buldyrev2010catastrophic,gomez2013diffusion,sahneh2013generalized,kivela2014multilayer,gomez2015layer}. The interactions between layers (networks) may yield the outcomes beyond what isolated layers can capture \cite{bauch2013social}. Granell $\emph{et}$ $\emph{al.}$ \cite{2013dynamical} proposed an unaware-aware-unaware (UAU) susceptible-infected-susceptible (SIS) dynamics in a multiplex network, which is composed of a virtual contact network and a physical contact network, to investigate the interplay between the spreading of awareness and epidemic. Different from the assumption that both the awareness and epidemic spreading processes have the same dynamics in \cite{2013dynamical}, Guo $\emph{et}$ $\emph{al.}$ \cite{guo2015two} introduced a threshold model to describe the awareness cascading phenomenon of human awareness. However, they both assumed that the aware individuals are completely immune to infection, that is, each individual chooses vaccination in response to information. In reality, behavioral adoption or response, especially vaccination, is a complex process \cite{young2011dynamics,centola2010spread,bauch2004vaccination}. On the one hand, vaccination is regarded as one of the most effective and protective behaviors (strategies) against virus propagation \cite{fu2011imitation,reluga2006evolving,xia2014belief}. On the other hand, vaccination usually comes with some cost, and the decision of an individual on vaccination depends on not only his trade-off to the cost, but also the strategies of other individuals. Vaccination presents a social dilemma since a self-interested individual expects to get benefit from the vaccinating behavior of others \cite{bauch2003group,galvani2007long,hilbe2014cooperation,zhang2014effects}. Therefore, understanding the relation between information dissemination and vaccination behavior is critical to epidemic control.

In this study, we construct an evolutionary vaccination game in a multiplex network which is composed of an information layer and a contact layer, and explore the role of information dissemination on vaccination. In order to reflect the reactions of individuals to risk information, we introduce an alert state (A) with information attributes into the SIS model, and propose a two-layer coupled SAIS model to describe the spreading process. We assume that the alert individuals are less likely to be infected than the susceptible individuals. We find that different behavioral responses to information have different impacts on epidemic spread. Moreover, we study the factors to affect vaccination equilibrium and social cost with numerical simulations.

The rest of this paper comes as follows. Section \ref{sec_2} formulates the problem of this paper. Section \ref{sec_3} presents a two-layer coupled SAIS model, where we analyse the epidemic dynamics theoretically and numerically. In section \ref{sec_4}, we introduce an evolutionary vaccination game in a multiplex network. Section \ref{sec_5} presents the vaccination performance against epidemic propagation and the role of information. Section \ref{sec_6} concludes the whole paper.

\section{\label{sec_2}Problem Formulation}

Vaccination is an effective and preventive strategy against epidemic propagation. When information and epidemic spread simultaneously, those susceptible individuals may get to know the epidemic status by receiving risk information from their infected neighbors. The qualitative analysis on the impact of information dissemination on vaccination cannot reach a unified conclusion. Consequently, Xia and Liu \cite{xia2014belief} proposed a belief-based model to study the impact factors of individual vaccination decisions. However, the costs of individuals related to the social cost were not taken into account. Note that it is generally assumed that individuals are self-interested to minimize their own costs instead of the social cost, which is one of the most important optimization subjects in epidemic control. An individual's vaccination decision depends on not only his trade-off on the cost but also the strategies of other individuals, since vaccination contributes to herd immunity, protecting individuals without vaccination from being infected. The correlation between information dissemination and vaccination determines the vaccination density that affects the social cost. Therefore, in this study, we propose an evolutionary vaccination game by integrating an information-epidemic spreading process into the process of strategic selection and interaction, and explore how information dissemination influences vaccination. Moreover, we show the social cost and infection density as a function of vaccination density to explore the role of vaccination in epidemic control, and study the impact of strength coefficient and individual sensitivity on the vaccination  equilibrium with numerical simulations.

In order to study the impact of information dissemination on vaccination, we assume that the vaccinated individuals are fully protected, regardless of the influence of vaccine efficiency \cite{cardillo2013evolutionary,steinegger2018interplay}. Taking into account the regularity of seasonal diseases and effectiveness of vaccination, we assume that individuals who are prone to immunization will be vaccinated before the outbreak of disease. The evolutionary vaccination game model (as illustrated in Fig. 1) includes two stages: the decision-making stage (stage 1) and the spreading stage (stage 2). During stage 2 (epidemic season) the epidemic and information (of the epidemic status) propagate simultaneously at the corresponding layer. Each node (individual) unilaterally decides whether to get vaccinated during stage 1, which is modeled by a vaccination game that occurs before the start of stage 2. A vaccinated individual will not be infected and no longer gets involved in the next epidemic season, while the unvaccinated individuals have a risk of being infected. When the epidemic process reaches a steady state, each individual will adjust his decision-making with respect to vaccination for the next epidemic season.

\begin{figure}[!htbp]
\centering
\includegraphics[totalheight=1.5in]{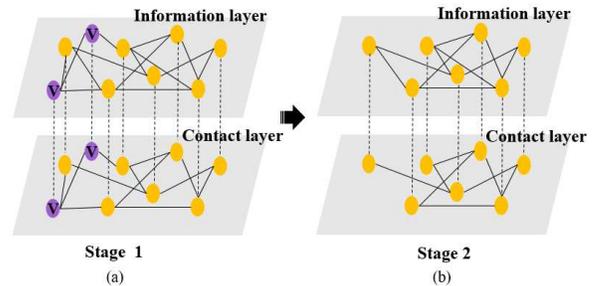}
\caption{A schematic stage illustration of the evolutionary vaccination game model in a multiplex network. This network includes two layers, the information layer corresponds to a network where risk information spreads, and the epidemic propagates on the contact layer. Individuals labeled by V choose to be vaccinated at stage 1. At stage 2, the vaccinated individuals are no longer involved in the epidemiological process which is described by the two-layer coupled SAIS model, while the unvaccinated individuals are at risk of being infected.  }
\end{figure}

\section{\label{sec_3}The SAIS Seasonal Epidemics without Vaccination}

\begin{figure*}[!htbp]
\centering
\includegraphics[totalheight=3in]{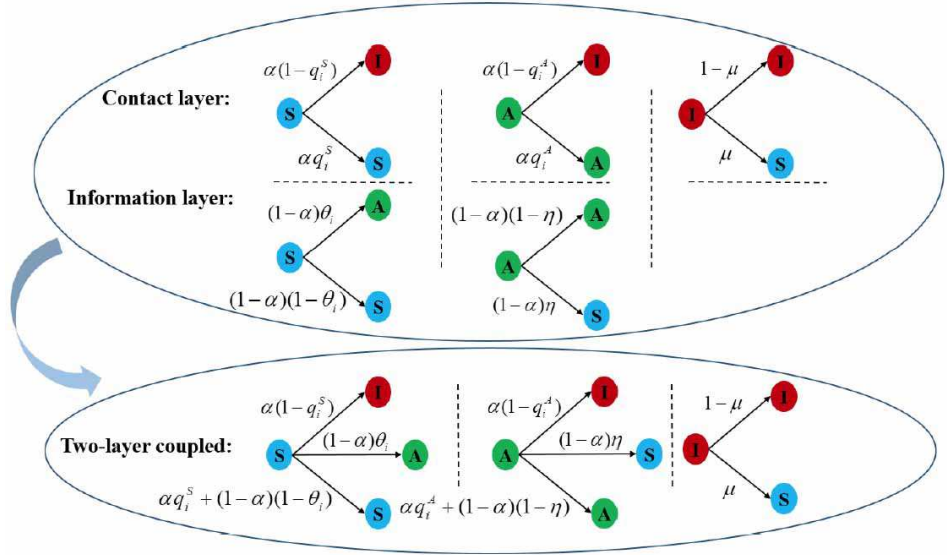}
\caption{Transition probability diagrams for susceptible (S), alert (A) and infected (I) states of the coupled SAIS propagation dynamics. Both susceptible and alert nodes can be infected by their infected neighbors, and the difference is that the alert nodes have been informed. $q_i^S$ denotes the transition probability for susceptible individual $i$ not being infected by the neighbours, $q_i^A$ denotes the transition probability for alert individual $i$ not being infected by the neighbours, $\theta_i$ denotes the transition probability for susceptible individual $i$ being informed by the infected neighbours, $\mu$ denotes the transition probability from the infected to the susceptible states, $\eta$ denotes the transition probability from the alert to the susceptible states, and $\alpha$ denotes the strength factor.}
\end{figure*}

A two-layer multiplex network with different network topologies are illustrated in Fig. 1. The information layer and contact layer have the same number of nodes, and the size is $N$. Each node (individual) in one layer has its counterpart node in another layer. There are three possible states for each node, susceptible (S), alert (A) or infected (I). In the information layer, a susceptible node can perceive the risk information of virus from its infected neighbors,  and convert to an alert node. Without loss of generality, we assume that an alert node may ignore or not care about this risk information, and becomes a susceptible node with rate $\eta$. In the contact layer, an infected node infects its susceptible and alert neighbours with infection rates $\beta$ and $\beta_A$, respectively, where $\beta_A=\xi\beta$. Taking into account the complexity of individual decision-making behaviour in reality, we assume that the alert state is different from the immune state, which means $0<\xi\leq1$. Each infected node recovers with rate $\mu$.

Let $p_i^S(t)$, $p_i^A(t)$ and $p_i^I(t)$ denote the probabilities for node $i$ of being susceptible, alert and infected at time $t$, respectively. We assume that each node has the same sensitivity $\lambda$ to the risk information. For susceptible node $i$ with degree $k_i$, the probability of being an alert node is denoted by $\theta_i=\frac{\lambda}{k_i}\sum_{j=1}^N a_{ji}p_j^I$, where $a_{ji}$ is the element of adjacency matrix $\mathcal{A}$ of the information layer, and $a_{ji}=1$ if there is a link between nodes $i$ and $j$. We denote the transition probability for node $i$ not being infected by the neighbours as $q_i^S(t)$ if $i$ is a susceptible node, or as $q_i^A(t)$ if $i$ is an alert node. The element of adjacency matrix $\mathcal{B}$ of the contact layer is denoted by $b_{ji}$, and we have

\begin{equation}\label{9}
\left\{
\begin{array}{l}
q_i^A(t)=\prod_{j=1}^N (1-b_{ji}p_j^I(t)\beta_A) \\
q_i^S(t)=\prod_{j=1}^N (1-b_{ji}p_j^I(t)\beta).\\
\end{array}
\right.
\end{equation}

We define the strength coefficient $\alpha$ as the tendency and intensity of information dissemination. The transition probability diagrams for three states of the coupled SAIS propagation dynamics are illustrated in Fig. 2. For instance, the probability that susceptible node $i$ remains susceptible at each time step is denoted by $\alpha q_i^S+(1-\alpha)(1-\theta_i)$. Specifically, $\alpha=1$ corresponds to the case in a single-layer network, i.e., only the epidemic propagates in the contact layer. The information and epidemic spreading processes coexist when $0<\alpha<1$.

The continuous time Markov approach can provide an exact description of the actual epidemic spreading, however, the infinitesimal generator $Q_{2^N\times2^N}$ \cite{van2009virus} is difficult to obtain, especially for large scale networks \cite{wang2017unification}, since the Markov chain contains $2^N$ states. Therefore, we utilize the microscopic Markov chain approach \cite{2013dynamical} to explore the probability evolution of different states for node $i$ as below:

\begin{equation}\label{9}
\left\{
\begin{array}{l}
p_i^S(t+1)=[\alpha q_i^S(t)+(1-\theta_i(t))(1-\alpha)]p_i^S(t)+\mu p_i^I(t)\\
~~~~~~~~~~~~~~~~~~+\eta(1-\alpha)p_i^A(t)\\
p_i^A(t+1)=[\alpha q_i^A(t)+(1-\eta)(1-\alpha)]p_i^A(t)\\
~~~~~~~~~~~~~~~~~~+(1-\alpha)\theta_i(t)p_i^S(t)\\
p_i^I(t+1)=\alpha(1-q_i^S(t))p_i^S(t)+\alpha(1-q_i^A(t))p_i^A(t)\\
~~~~~~~~~~~~~~~~~~+(1-\mu)p_i^I(t). \\
\end{array}
\right.
\end{equation}

When $p_i^I(t+1)=p_i^I(t)=p_i^I$, we have
\begin{equation}\label{9}
\left\{
\begin{array}{ccl}
p_i^S&=&\frac{(1-p_i^I)[\alpha(1-q_i^A)+\eta(1-\alpha)]}{\alpha(1-q_i^A)+(1-\alpha)(\eta+\theta_i)}\\
p_i^A&=&\frac{(1-\alpha)\theta_i(1-p_i^I)}{\alpha(1-q_i^A)+(1-\alpha)(\eta+\theta_i)}\\
\frac{\mu}{\alpha}p_i^I&=&(1-q_i^S)p_i^S+(1-q_i^A)p_i^A. \\
\end{array}
\right.
\end{equation}

Let $\beta_A=\xi\beta$, combining Eq. (3) with $p_i^S+p_i^A+p_i^I=1$, we obtain the infection probability of node $i$ in the stationary state,
\begin{equation}
p_i^I=\frac{M+\alpha\eta(1-\alpha)(1-q_i^S)}{M+\alpha\mu(1-q_i^A)+(1-\alpha)[\mu(\eta+\theta_i)+\alpha\eta(1-q_i^S)]},
\end{equation}
where $M=\alpha(1-q_i^A)[(1-\alpha)\theta_i+\alpha(1-q_i^S)]$. Thus, the infection density $\rho^I$ can be computed as
\begin{equation}
\rho^I=\frac{1}{N}\sum_{i=1}^Np_i^I.
\end{equation}

We compare $\rho^I$ obtained by Eq. (5) with the one obtained by Monte Carlo (MC) simulations to evaluate the analytical result. Simulations are performed in two-layer ER networks and two-layer BA scale-free networks, where the network size (each layer with the same size) and strength coefficient are taken into account to illustrate the applicability of the model. In a multiplex network with two layers, the topology of each layer is different. For instance, the contact layer of a two-layer BA scale-free network has a power-law degree distribution with exponent 3, and the information layer is the same network with some extra random links. Fig. 3 shows that the infection density $\rho^I$ increases with the increase of infection rate $\beta$. When $\beta>\beta_c$, the so-called epidemic threshold, the epidemic outbreaks and infection density $\rho^I>0$.

\begin{figure}
\subfigure[]{
\begin{minipage}[t]{0.45\linewidth}
\centering
\includegraphics[height=45mm,width=100mm]{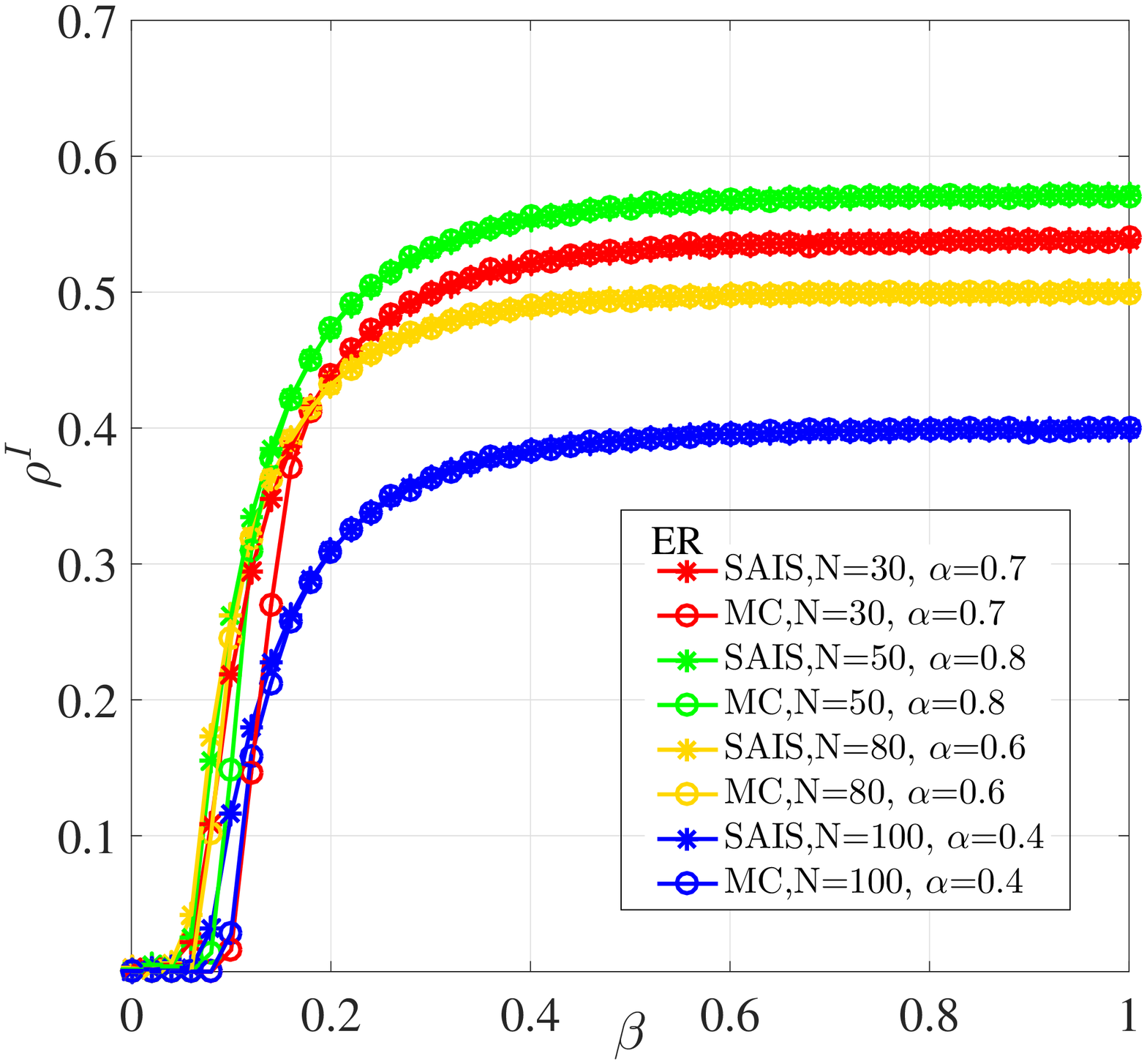}
\end{minipage}
}
\subfigure[]{
\begin{minipage}[t]{0.45\linewidth}
\centering
\includegraphics[height=45mm,width=100mm]{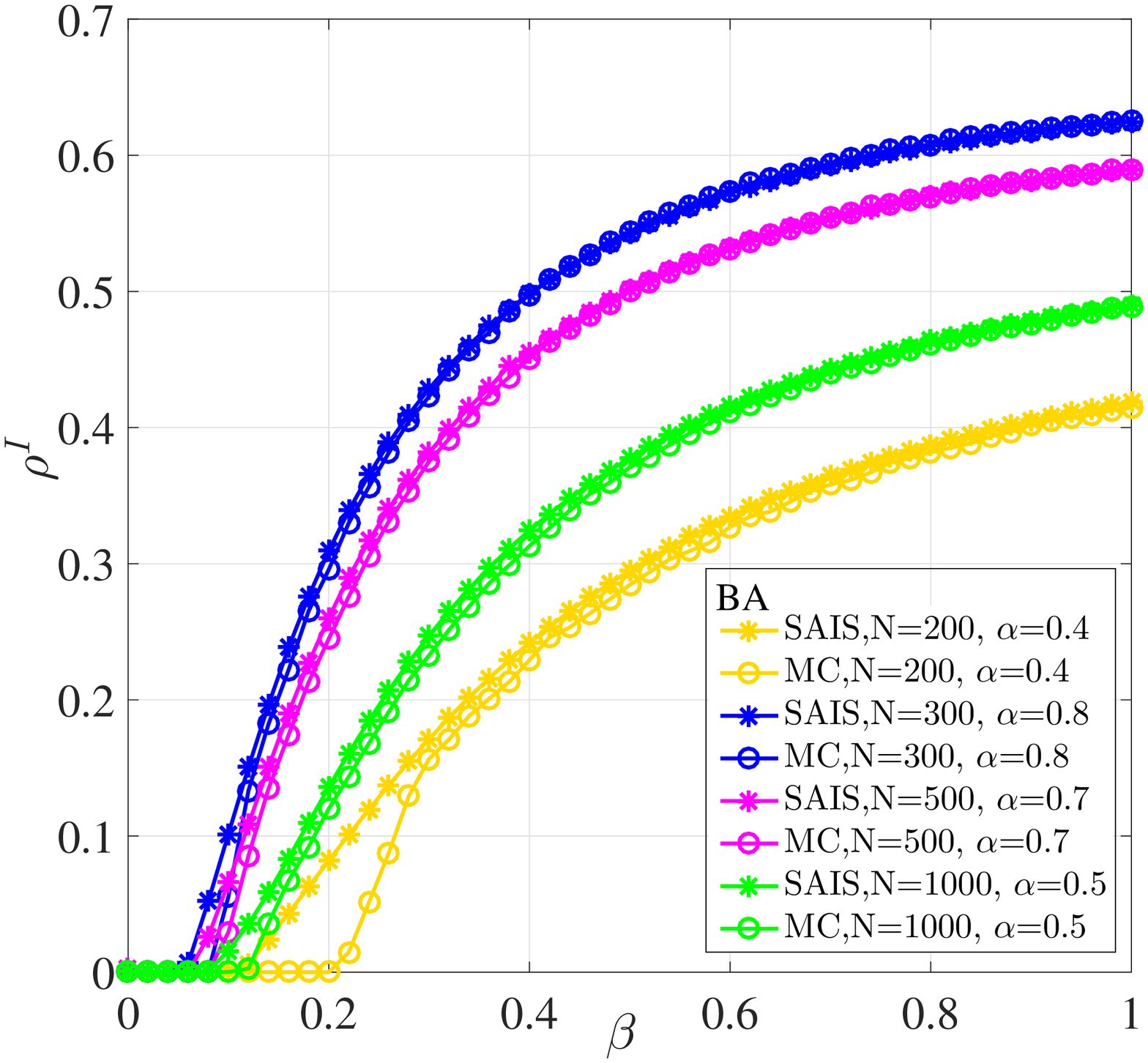}
\end{minipage}
}
\caption{The comparison of infection density $\rho^I$ obtained by the SAIS model and Monte Carlo simulations as a function of infection rate $\beta$ in (a) two-layer ER networks and (b) two-layer BA scale-free networks with exponent 3, respectively. The Monte Carlo simulations is averaged by 30 realizations. $\mu=0.6$, $\eta=0.1$, $\xi=0.5$.}
\end{figure}
When $\beta\to\beta_c$, the probability satisfies $0\leq p_i^I\ll1$, Eq. (2) can be further simplified as

\begin{equation}\label{9}
\left\{
\begin{array}{l}
q_i^A(t)=1-\xi\beta \sum_{j=1}^N b_{ji}p_j^I(t) \\
q_i^S(t)=1-\beta \sum_{j=1}^N b_{ji}p_j^I(t).\\
\end{array}
\right.
\end{equation}

Let $\phi_i=p_i^I$, combining Eq. (3) with Eq. (6) and omitting the second-order terms of $\phi$, we obtain
\begin{equation}
\frac{\mu}{\alpha\beta}\phi_i=(1-(1-\xi)p_i^A)\sum_{j=1}^N b_{ji}\phi_j.
\end{equation}

Considering that $\theta_i$ is proportional to the sum of $p_j^I$, we obtain $0\leq p_i^A\ll1$. Then, Eq. (7) can be reduced to
\begin{equation}
\sum_{j=1}^N[b_{ji}-\frac{\mu}{\alpha\beta}\epsilon_{ji}]\phi_j=0,
\end{equation}
where $\epsilon_{ji}$ is the element of the identify matrix. Eq. (8) has non-trivial solutions if and only if $\frac{\mu}{\alpha\beta}$ is the eigenvalue of adjacency matrix $\mathcal{B}$. Therefore, we obtain the epidemic threshold
\begin{equation}
\beta_c=\frac{\mu}{\alpha\Delta_{max}(\mathcal{B})},
\end{equation}
where $\Delta_{max}(\mathcal{B})$ is the largest eigenvalue of matrix $\mathcal{B}$. Obviously, the epidemic threshold $\beta_{c}$ depends on the structure of contact layer $\mathcal{B}$ and the strength coefficient $\alpha$.

\begin{figure}[!htbp]
\subfigure[]{
\begin{minipage}[t]{0.45\linewidth}
\centering
\includegraphics[height=45mm,width=100mm]{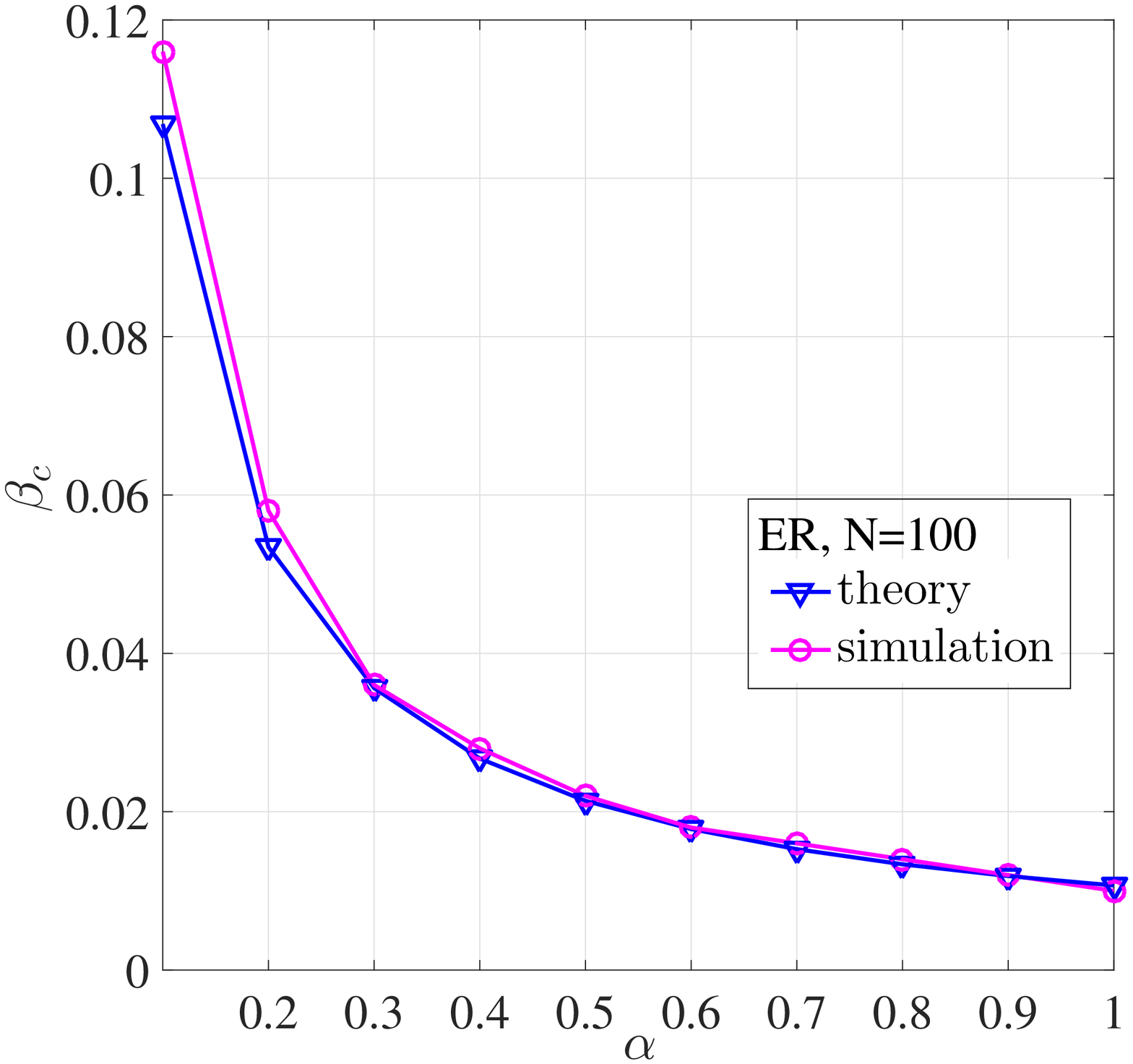}
\end{minipage}
}
\subfigure[]{
\begin{minipage}[t]{0.45\linewidth}
\centering
\includegraphics[height=45mm,width=100mm]{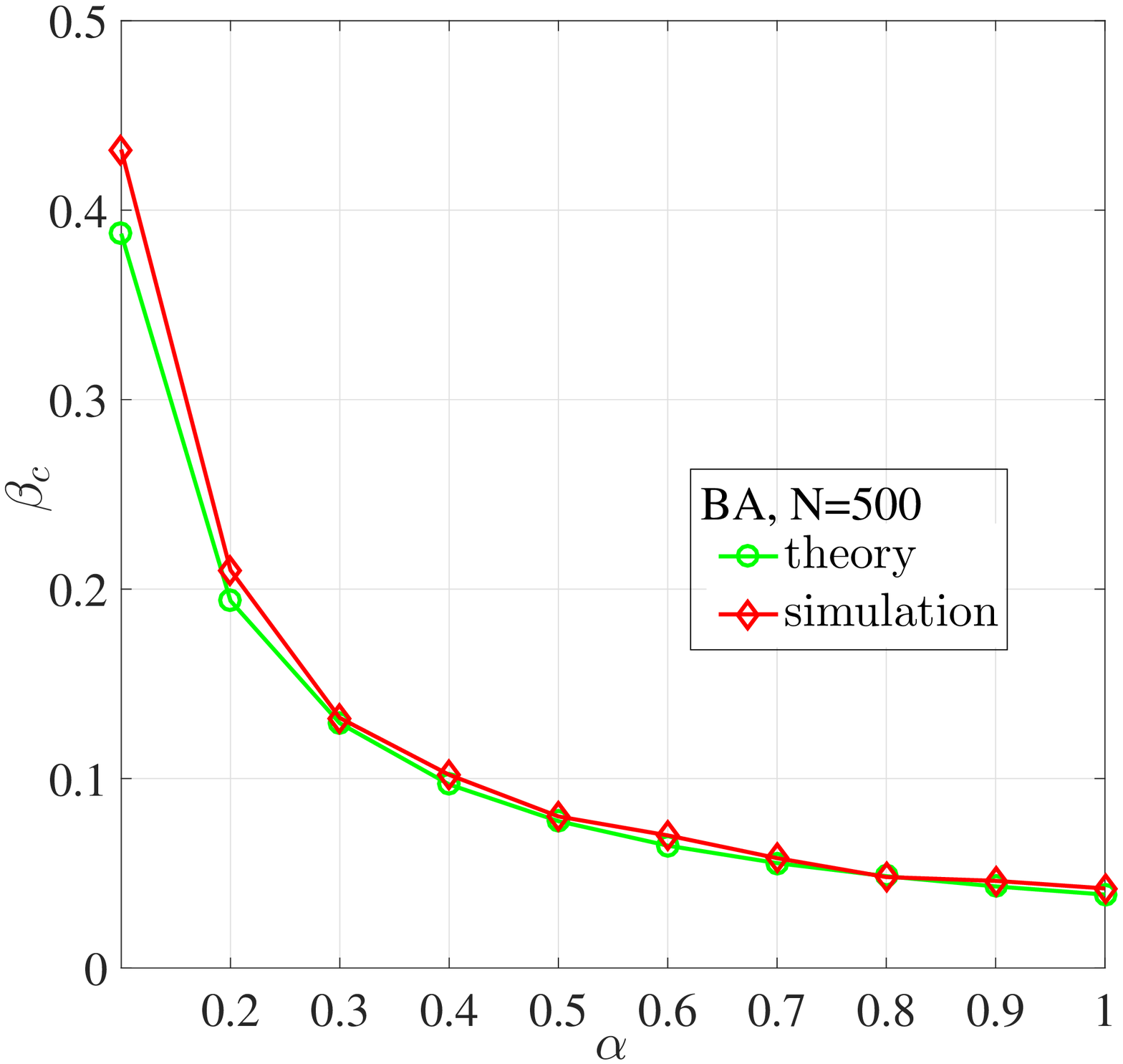}
\end{minipage}
}
\caption{Epidemic threshold $\beta_c$ as a function of strength coefficient $\alpha$ in (a) a two-layer ER network with $N=100$ in each layer and (b) a two-layer BA scale-free network with $N=500$ in each layer, respectively. $\mu=0.45$, $\eta=0.1$, $\xi=0.5$.}
\end{figure}

\begin{figure*}[!htbp]
\centering
\subfigure[~~~~~~]{
\begin{minipage}{0.3\linewidth}
\centering
\includegraphics[height=42mm,width=100mm]{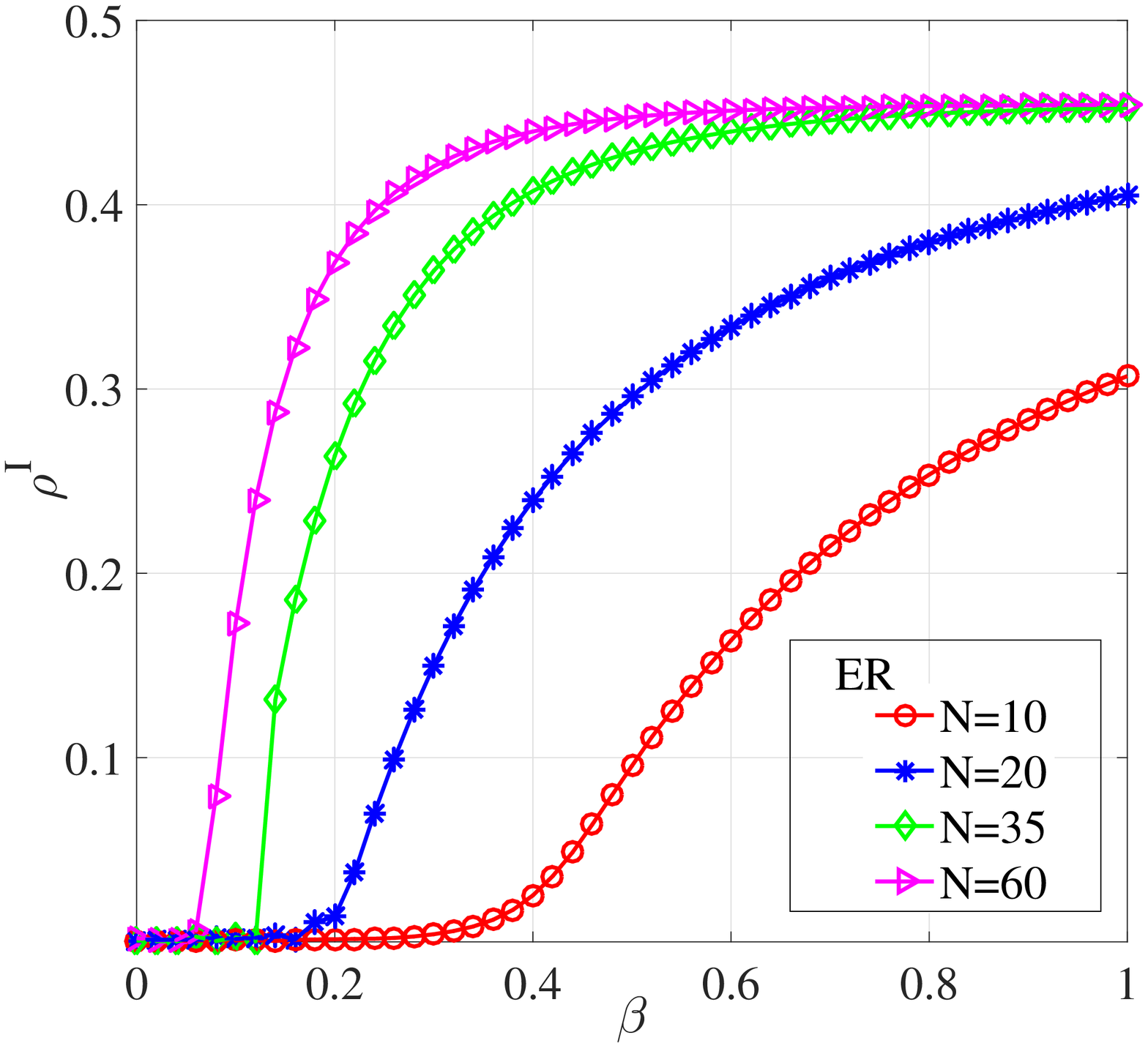}
\end{minipage}
}
\subfigure[~~~~~~]{
\begin{minipage}{0.3\linewidth}
\centering
\includegraphics[height=42mm,width=100mm]{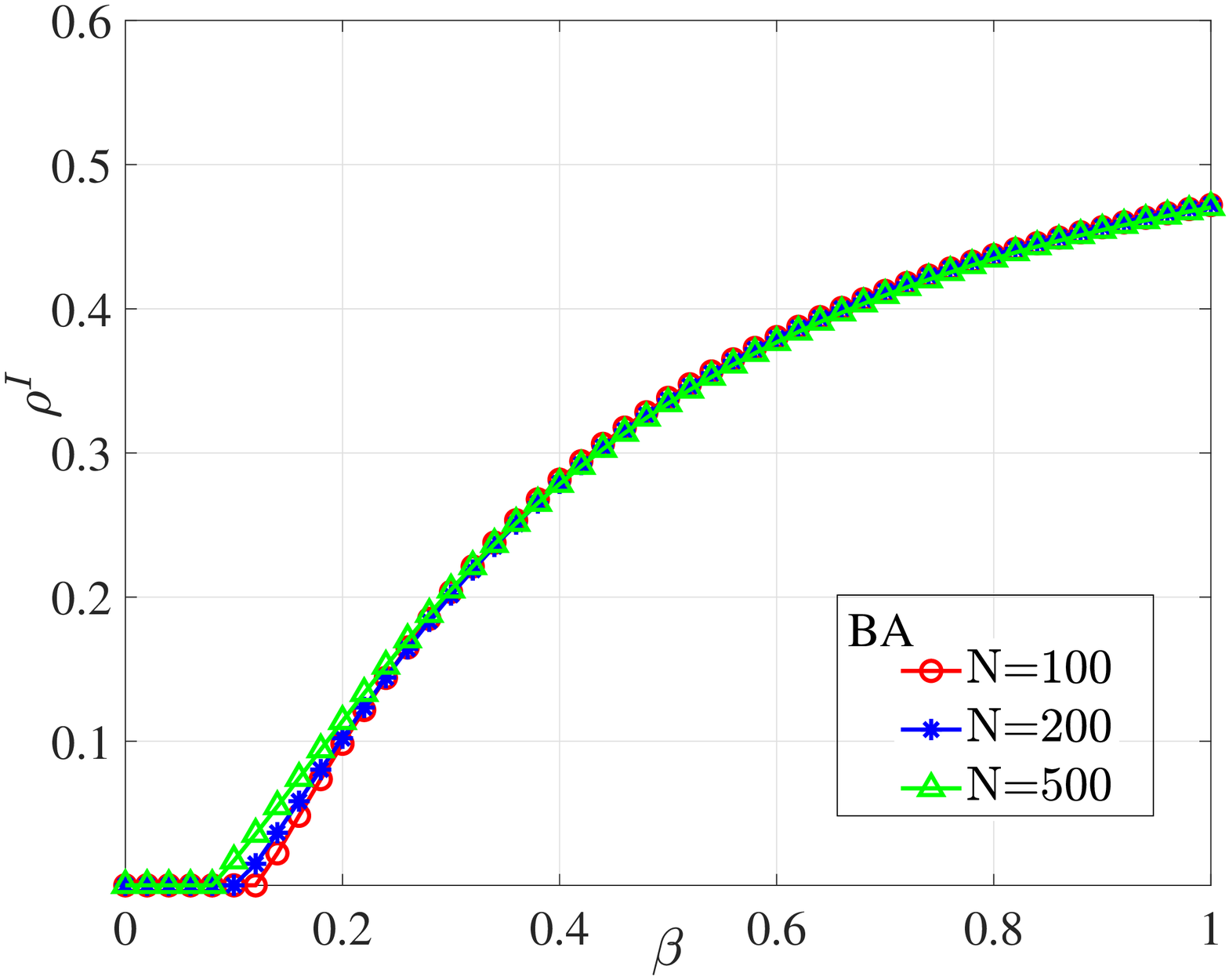}
\end{minipage}
}
\subfigure[~~~~~~]{
\begin{minipage}{0.3\linewidth}
\centering
\includegraphics[height=42mm,width=100mm]{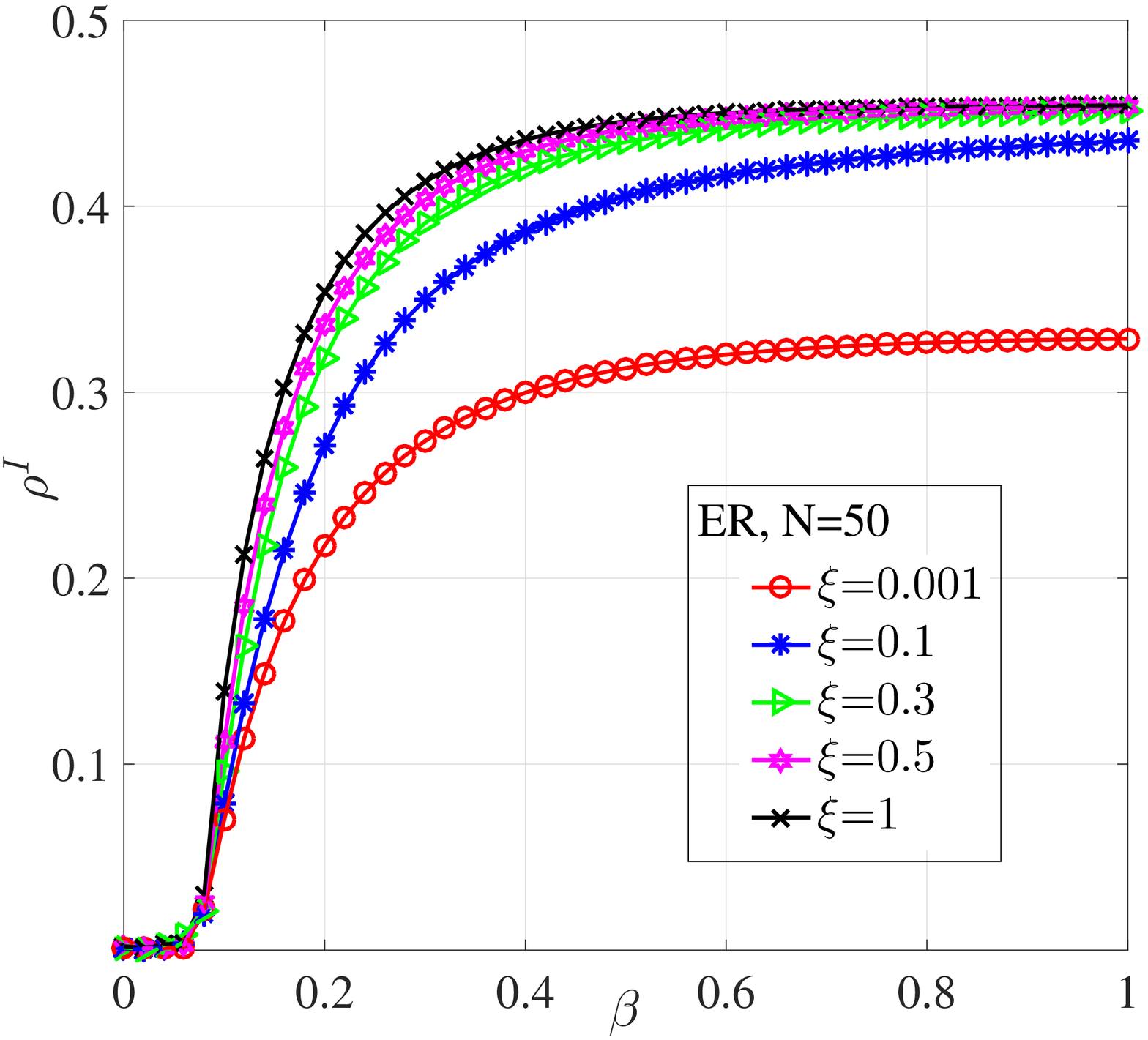}
\end{minipage}
}
\caption{Infection density $\rho^I$ as a function of infection rate $\beta$ in multiplex networks with different $N$, where each layer is (a) an ER network and (b) a BA scale-free network, respectively. The strength coefficient $\alpha=0.5$. (c) The impacts of coefficient $\xi$ on the infection density $\rho^I$ and epidemic threshold $\beta_c$.}
\end{figure*}

Fig. 4 illustrates the impacts of strength coefficient $\alpha$ on epidemic threshold $\beta_c$. We find that the epidemic threshold decreases with the increase of $\alpha$, regardless of the network topology. Moreover, the epidemic threshold calculated by Eq. (9) is in agreement with the one obtained by the MC simulations in a two-layer ER network and a two-layer BA scale-free network, respectively. In terms of the infection density and epidemic threshold, we conclude that there is a good agreement between the MMCA method and Monte Carlo simulations no matter what network sizes and strength coefficients are.

Besides, network size $N$ also influences infection density $\rho^I$ and epidemic threshold $\beta_c$. Fig. 5 shows that the epidemic threshold $\beta_c$ decreases with the increase of network size $N$. In terms of the infection density $\rho^I$, it depends on not only the network size $N$ but also the infection rate $\beta$. Meanwhile, different network structures display different results. For a two-layer ER network (see Fig. 5 (a)), when the network size and the infection rate exceed a certain value ($N>35, \beta>0.6$), both network size $N$ and infection rate $\beta$ will no longer affect the infection density $\rho^I$. For a two-layer BA scale-free network, network size $N$ affects the infection density $\rho^I$ only when infection rate $\beta<0.4$. When $\beta\ge0.4$, the infection density $\rho^I$ only depends on $\beta$.

In order to reveal the role of behavioral response of the alert individuals, we investigate the impact of coefficient $\xi$ on the propagation dynamics. As illustrated in Fig. 5 (c), $\xi$ does not affect the epidemic threshold $\beta_{c}$, but affects the infection density $\rho^I$. Moreover, we find that the impact of $\xi$ on the epidemic threshold $\beta_c$ is various in different ranges of $\xi$. When $\xi>0.3$, the changes of $\xi$ will have no effect on the epidemic threshold $\beta_{c}$. The infection density is greatly reduced when $\xi \to0$, where all the alert individuals are immune to the infection, as the assumption in \cite{2013dynamical}. However, individuals with risk information do not necessarily choose vaccination in reality.

\section{\label{sec_4}Evolutionary Vaccination Game}

For a network with two layers, although different layers represent social interactions at different contexts, the strategic choices of an individual in one layer may affect that in the other layer. For simplicity, we assume that each node and its counterpart node have the same strategy during the same round.

The decision of an individual may be affected by many factors due to the infection interactions. We consider that there exists a communication cost $T$ due to the risk information diffuses in the information layer. In order to promote vaccination, we assume that unvaccinated individuals have communication costs while vaccinated ones do not. An individual choosing vaccination has the vaccination cost $C$, and vaccination is completely effective so that a vaccinated individual does not participate in the information-epidemic process. In addition to the communication cost, a non-vaccinated individual being infected has the infection cost $H$. Considering that an individual's strategy functions in both layers, we assume that the cost of an individual is determined by the sumed costs of two layers. The cost of individual $i$ in a two-layer coupled network is therefore denoted by
\begin{equation}
U_i=Cm_s+[Hv_i+(1-\alpha)T](1-m_s)
\end{equation}
where $m_s=1$ if individual $i$ chooses vaccination, otherwise, $m_s=0$, and $v_i=1$ or $0$ indicates whether individual $i$ is infected or not. Without loss of generality, we assume that cost $C$ of a vaccinated individual is less than cost $H$. Meanwhile, an unvaccinated individual who benefits from his vaccinated neighbours may not be infected, resulting in a social dilemma. Hence, we have the following condition
\begin{equation}
(1-\alpha)T<C<H+(1-\alpha)T
\end{equation}

We define the proportion of individuals who choose vaccination as vaccination density, denoted by $x$. The infection rate becomes $(1-x)\beta$ during the epidemic season \cite{anderson1992infectious}. The average payoffs of vaccinated individuals $P_v$ and unvaccinated individuals $P_{uv}$ are as follows,
\begin{equation}\label{9}
\left\{
\begin{array}{ccl}
P_v&=&-C\\
P_{uv}&=&-[f(x)( H+(1-\alpha)T)+(1-f(x))(1-\alpha)T],
\end{array}
\right.
\end{equation}
where $f(x)=\frac{\rho^I(x)}{1-x}$ is the ratio of the number of infected individuals to that of unvaccinated individuals. The infection density
\begin{equation}
\rho^I(x)=\frac{1}{N}\sum_{i=1}^Np_i^I(x),
\end{equation}
where $p_i^I(x)$ is the infection probability for node $i$ in the stationary state, which can be obtained in section \ref{sec_3}.  The social cost of a multiplex network with $N$ nodes is

\begin{equation}
\begin{array}{l}
E_{sc}=N[xC+[(H+(1-\alpha)T)f(x) \\
 ~~~~~~~~~+(1-\alpha)(1-f(x))T](1-x) \\
  ~~~~=N[(C-(1-\alpha)T)x+\rho^I(x)+(1-\alpha)T].
\end{array}
\end{equation}
The optimal value of vaccination density $x$ to minimize the social cost is herd immunity threshold $x_c$. We will discuss the factors which influence the social cost and herd immunity threshold $x_c$ in section \ref{sec_5}.

We study the vaccination dynamics and predict vaccination behavior of individuals through pairwise interactions in a two-layer coupled network. Once the spreading process in this season ends, each individual updates his strategy for the next epidemic season. We adopt the Fermi rule \cite{szabo1998evolutionary} for the strategy updating. At each round, individual $i$ randomly selects a neighbour $j$ in the information layer, compares their costs, and learns the strategy of individual $j$ with the following probability:
\begin{equation}
w_{(S_i\gets S_j)}=\frac{1}{1+\text{exp}^{[-k(U_j-U_i)]}},
\end{equation}
where $S_i$ and $S_j$ correspond to the strategies of individuals $i$ and $j$, respectively. Parameter $k$ represents the selection intensity, measuring how much the selection depends on the cost difference.

In homogeneous networks, we approximate the evolutionary dynamics with the replicator dynamics \cite{taylor1978evolutionary}, which is presented as $\dot{x}=x(P_v-\bar{P})$, where $\bar{P}=xP_v+(1-x)P_{uv}$ is the average payoff. Therefore,

\begin{equation}
\begin{array}{ccl}
\frac{\text{d}x}{\text{d}t}&=&x(P_v-\bar{P}) \\
  &=&x(1-x)[Hf(x)+(1-\alpha)T-C],\\
\end{array}
\end{equation}
where vaccination cost $C\in[0,1)$. For simplicity, let $H=1$,  Eq. (16) can be simplified as
\begin{equation}
\begin{array}{l}
\frac{\text{d}x}{\text{d}t}=x(1-x)[f(x)-C+(1-\alpha)T]. \\

\end{array}
\end{equation}

The equilibria of Eq. (17) are given by $x=0$, $x=1$, or interior equilibrium $\hat{x} \in (0,1)$ satisfying
\begin{equation}
f(\hat{x})- C+(1-\alpha)T=0.
\end{equation}
Note that $f(x)$ decreases with the increase of $x$, until $x$ reaches the herd immunity threshold $x_c$. If $\frac{C}{T}>1-\alpha$, when $x\in(0,\hat{x})$, we have $ f(x)- C+(1-\alpha)T>0$, similarly, when $x\in(\hat{x},1)$, we have $ f(x)- C+(1-\alpha)T<0$. Based on the equilibrium stability criterion \cite{nowak2006evolutionary}, we conclude that the interior equilibrium $\hat{x}$ is stable, given the equation $f(\hat{x})=C-(1-\alpha)T$ is satisfied. The equilibrium $x=0$ is stable, if $C>\rho^I(0)$ and $ f(x)-C+(1-\alpha)T<0$, when $x=0$ and $f(0)=\rho^I(0)$. That is, in a homogeneous multiplex network, the costs of vaccination and communication are $C$ and $T$, where $C, T\in[0,1)$, respectively. The stable vaccination equilibrium $x^*=\hat{x}$, satisfying $f(x^*)=C-(1-\alpha)T$, is an evolutionary stable strategy (ESS), when $\frac{C}{T}>1-\alpha$; the ESS is $x^*=0$, when $C>\rho^I(0)+(1-\alpha)T$.

\section{\label{sec_5}Vaccination performance and role analysis of information}

\begin{figure}[!htbp]
\centering
\includegraphics[height=100mm,width=85mm]{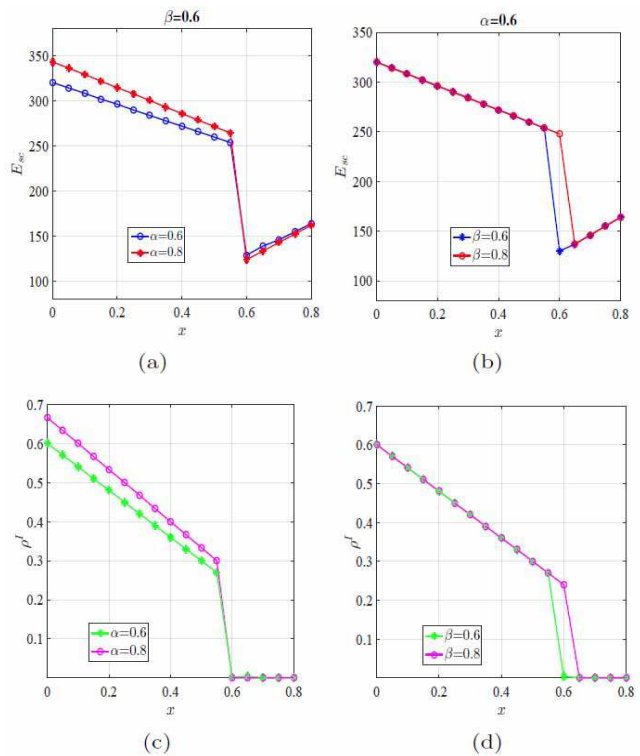}
\caption{Social cost (a) and infection density (c) as a function of vaccination density $x$ with different strength coefficients $\alpha$. Social cost (b) and infection density (d) as a function of vaccination density $x$ with different infection rates $\beta$. $N=500$, $\mu=0.4$, $C=0.4$, $T=0.1$ and $\xi=0.5$.  }
\end{figure}

\begin{figure*}[!htbp]
\centering
\subfigure[~~~~~~]{
\begin{minipage}{0.3\linewidth}
\centering
\includegraphics[height=42mm,width=100mm]{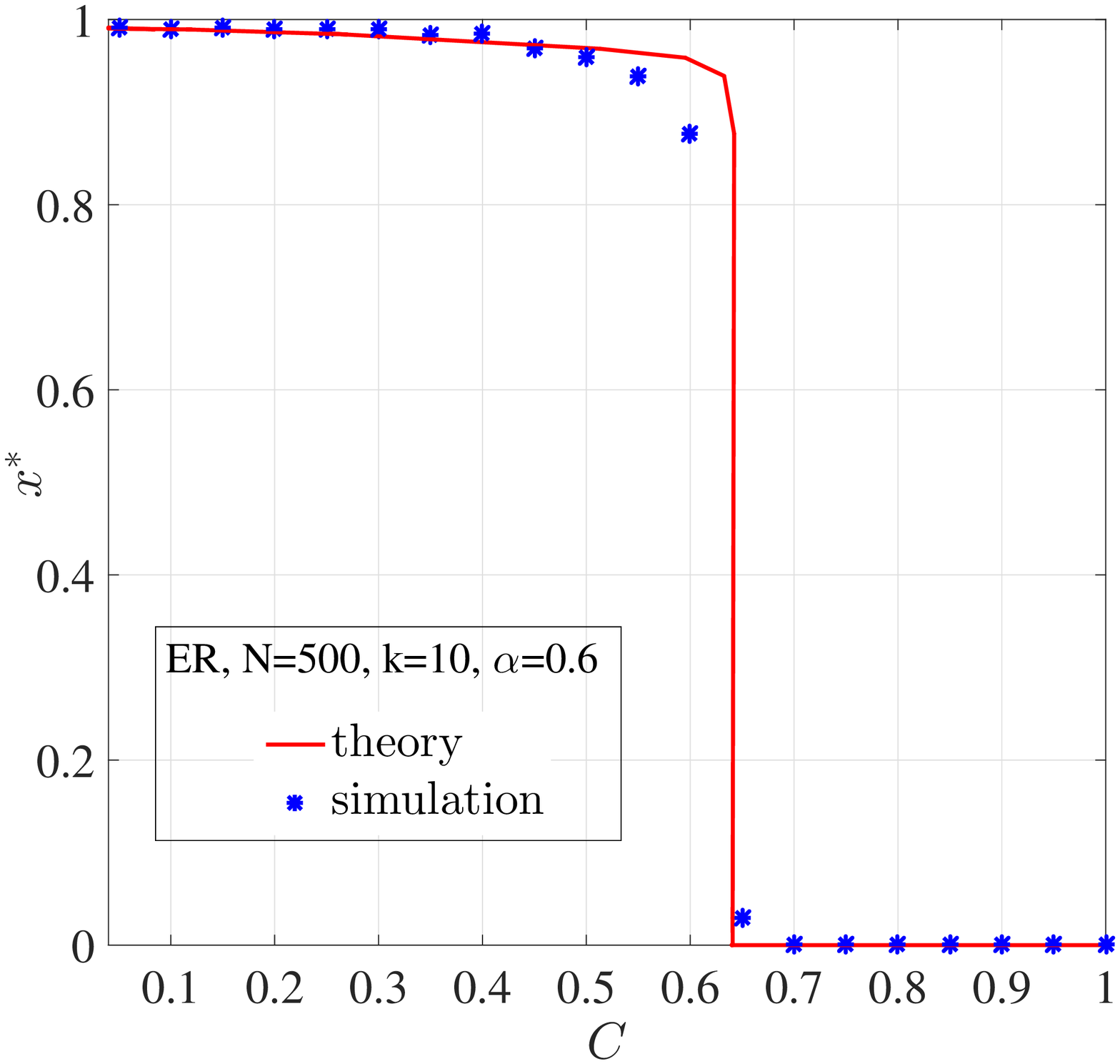}
\end{minipage}
}
\subfigure[~~~~~~]{
\begin{minipage}{0.3\linewidth}
\centering
\includegraphics[height=42mm,width=100mm]{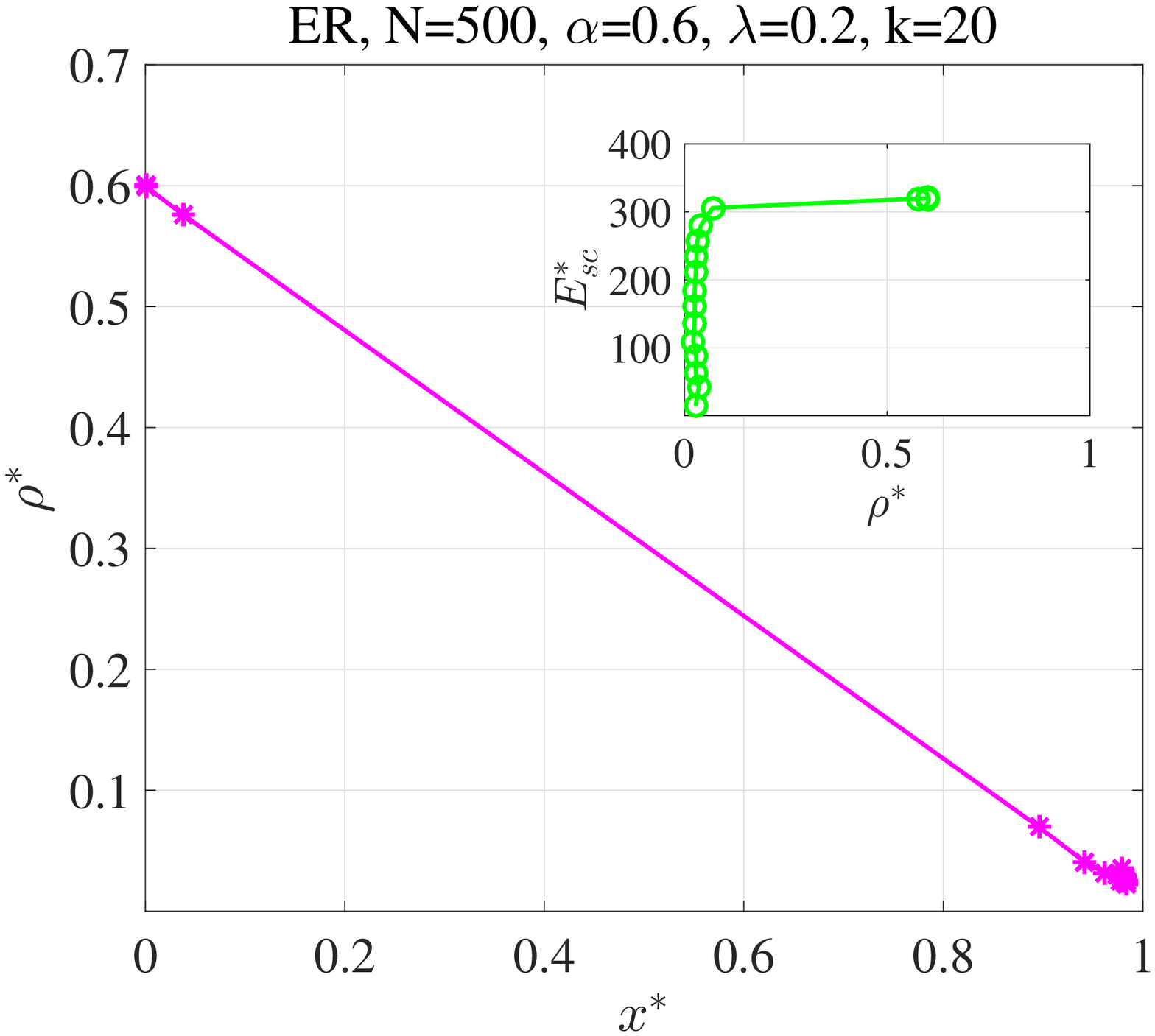}
\end{minipage}
}
\subfigure[~~~~~~]{
\begin{minipage}{0.3\linewidth}
\centering
\includegraphics[height=42mm,width=100mm]{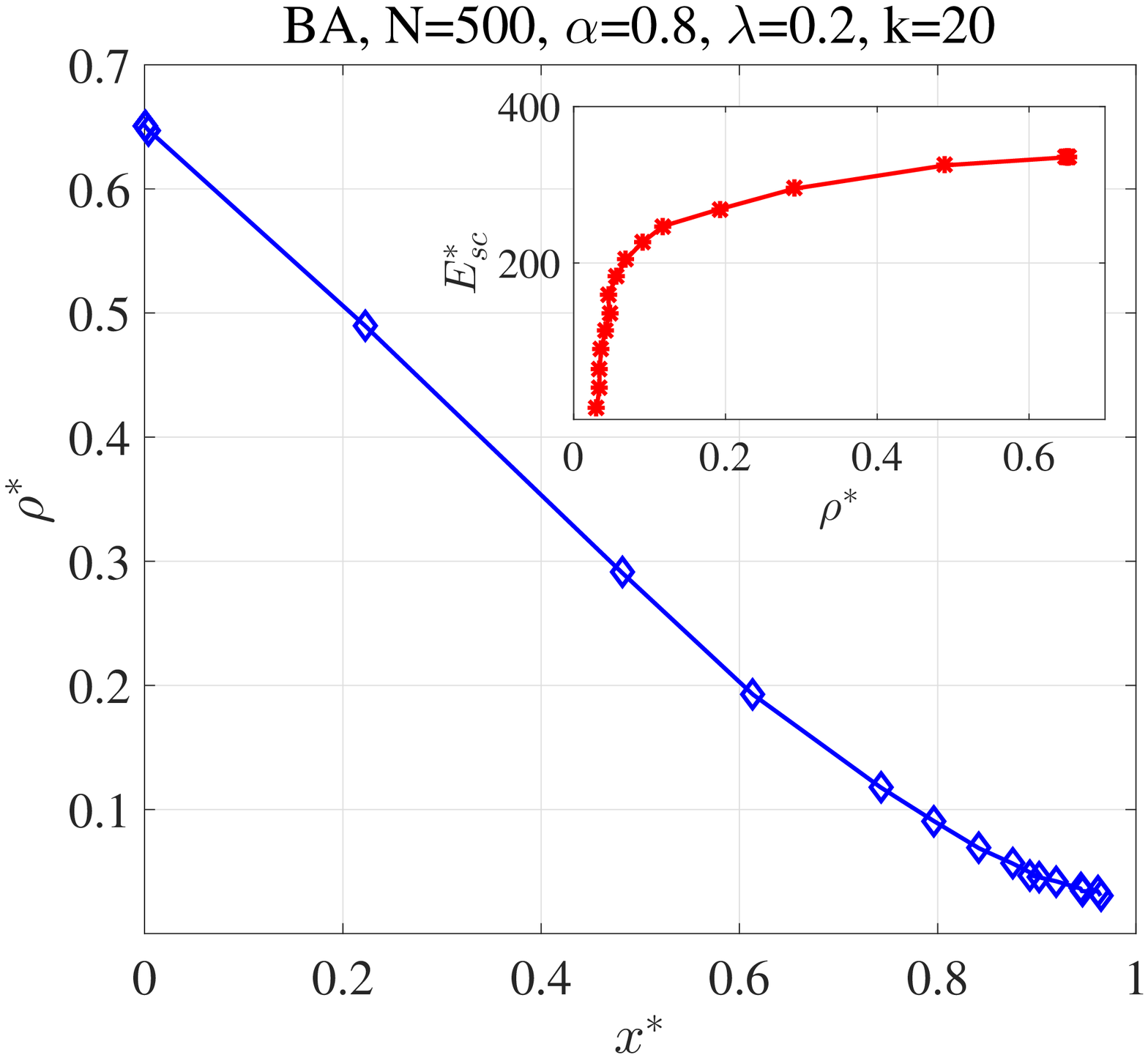}
\end{minipage}
}
\caption{(a) The relationship between the vaccination equilibrium $x^*$ and vaccination cost $C$. The solid line is numerically determined by Eq. (18). The final infection density $\rho^*$ as a function of the vaccination equilibrium $x^*$ in a two-layer ER network (b) and a two-layer BA scale-free network (c), respectively. Insets show the social cost $E_{sc}^*$ as a function of final infection density $\rho^*$. $\mu=0.4$, $\beta=0.8$, $T=0.1$ and $\xi=0.5$.}
\end{figure*}
\begin{figure*}[!htbp]
\centering
\subfigure[~~~~~~]{
\begin{minipage}{0.3\linewidth}
\centering
\includegraphics[height=42mm,width=100mm]{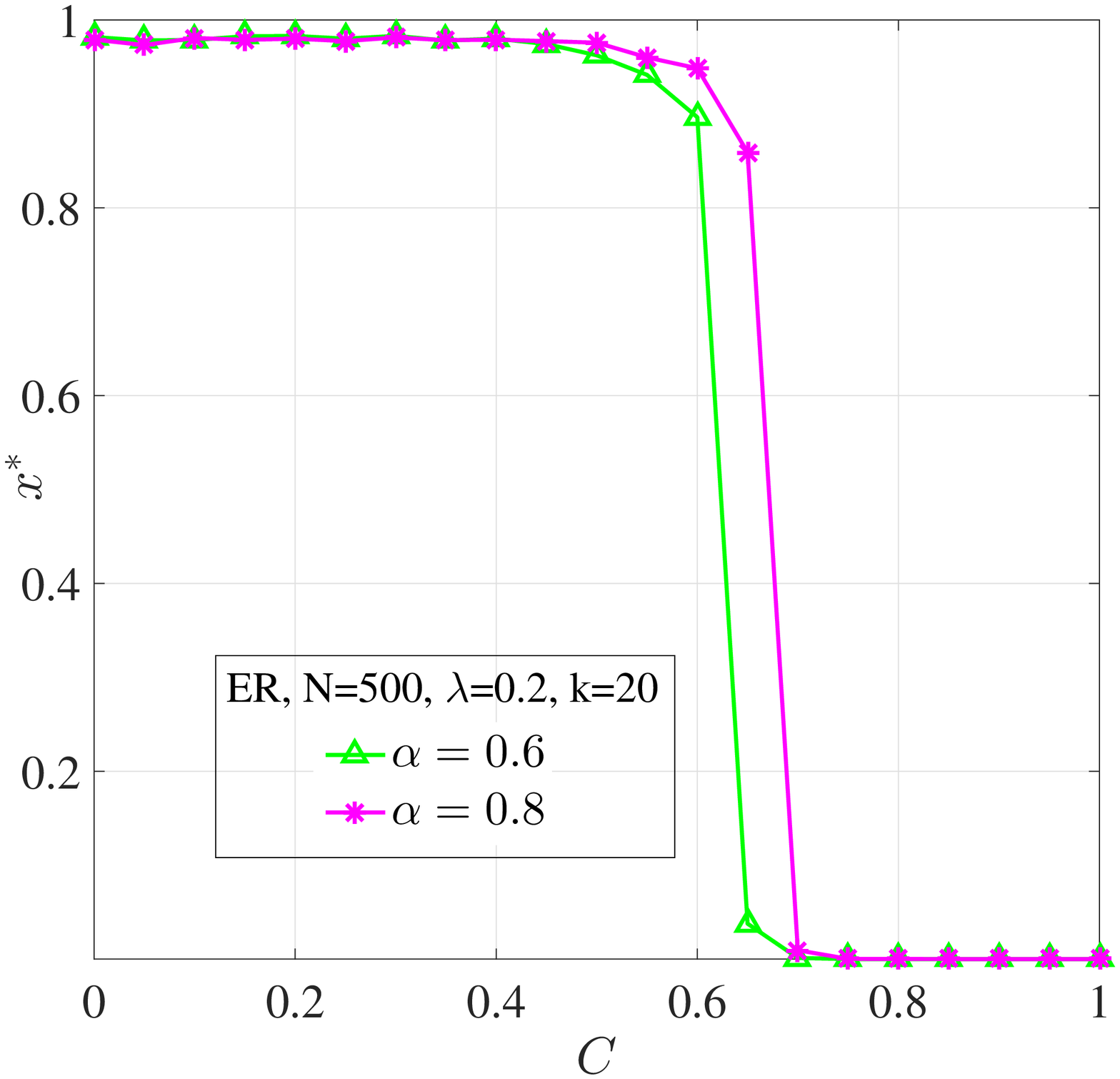}
\end{minipage}
}
\subfigure[~~~~~~]{
\begin{minipage}{0.3\linewidth}
\centering
\includegraphics[height=42mm,width=100mm]{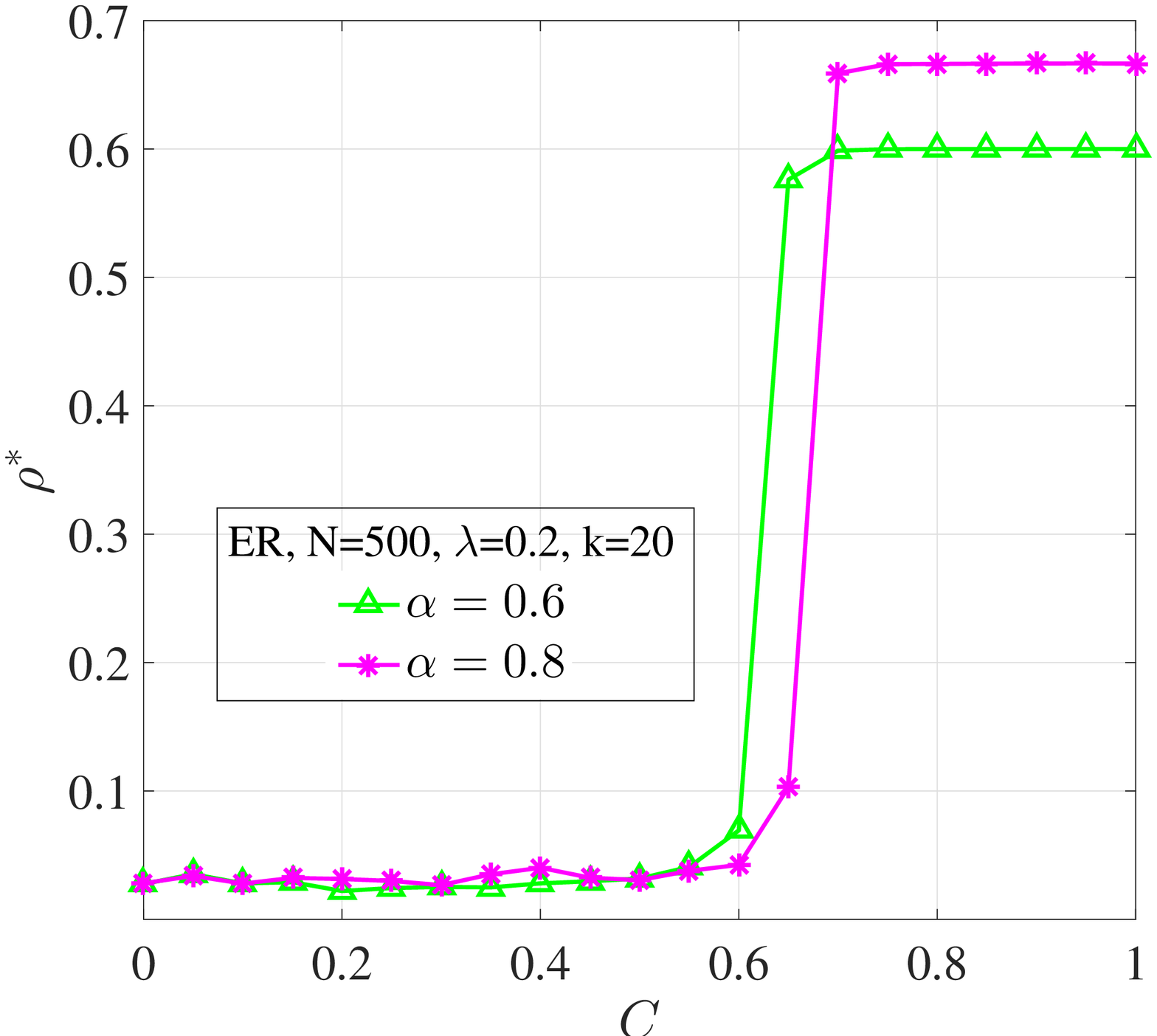}
\end{minipage}
}
\subfigure[~~~~~~]{
\begin{minipage}{0.3\linewidth}
\centering
\includegraphics[height=42mm,width=100mm]{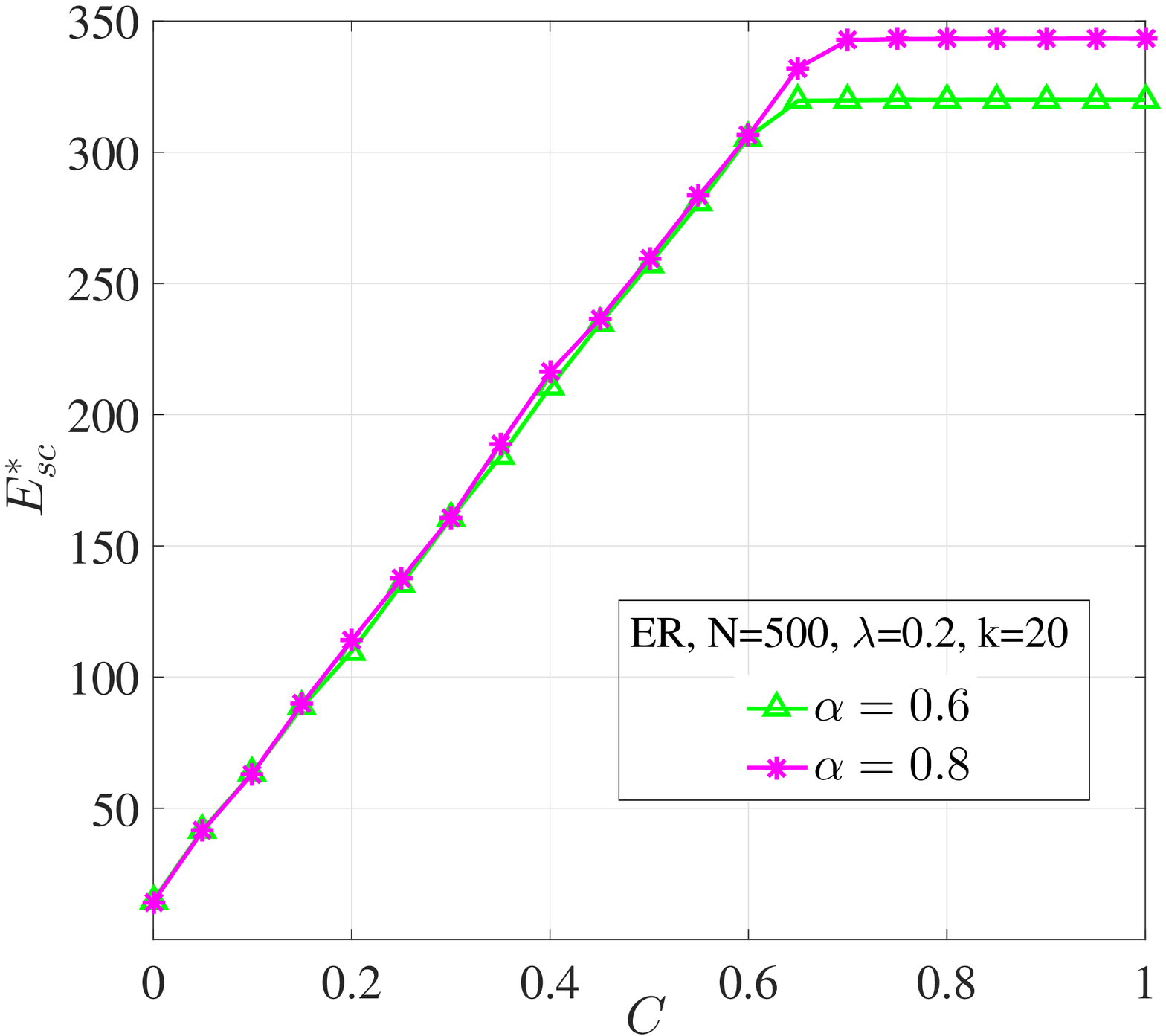}
\end{minipage}
}
\caption{For different strength coefficients, (a) vaccination equilibrium $x^*$, (b) final infection density $\rho^*$ and (c) social cost $E_{sc}^*$ as a function of vaccination cost $C$ in a two-layer ER network. $\mu=0.4$, $\beta=0.8$, $T=0.1$ and $\xi=0.5$.}
\end{figure*}

\begin{figure*}[!htbp]
\centering
\subfigure[~~~~~~]{
\begin{minipage}{0.3\linewidth}
\centering
\includegraphics[height=42mm,width=100mm]{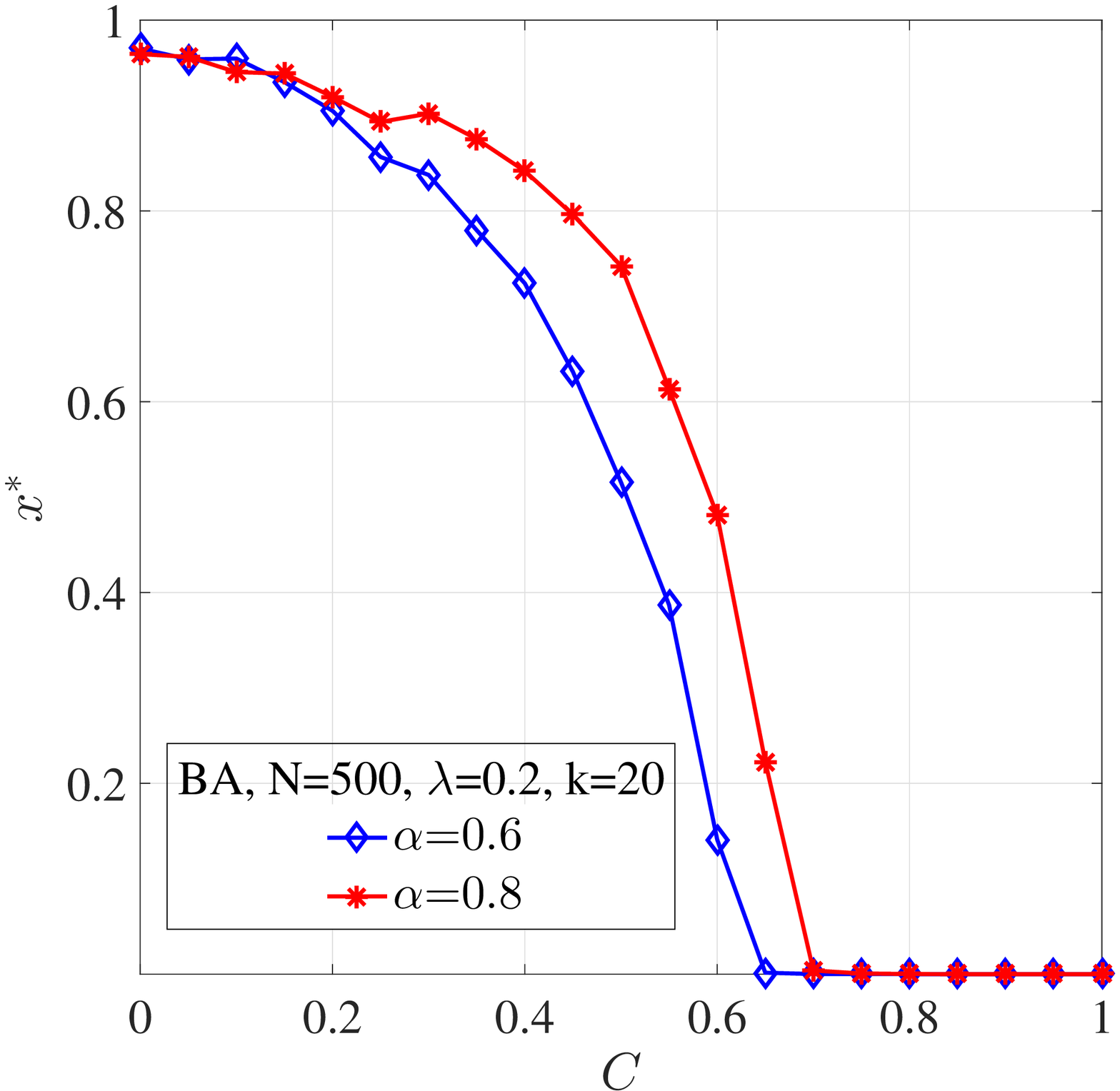}
\end{minipage}
}
\subfigure[~~~~~~]{
\begin{minipage}{0.3\linewidth}
\centering
\includegraphics[height=42mm,width=100mm]{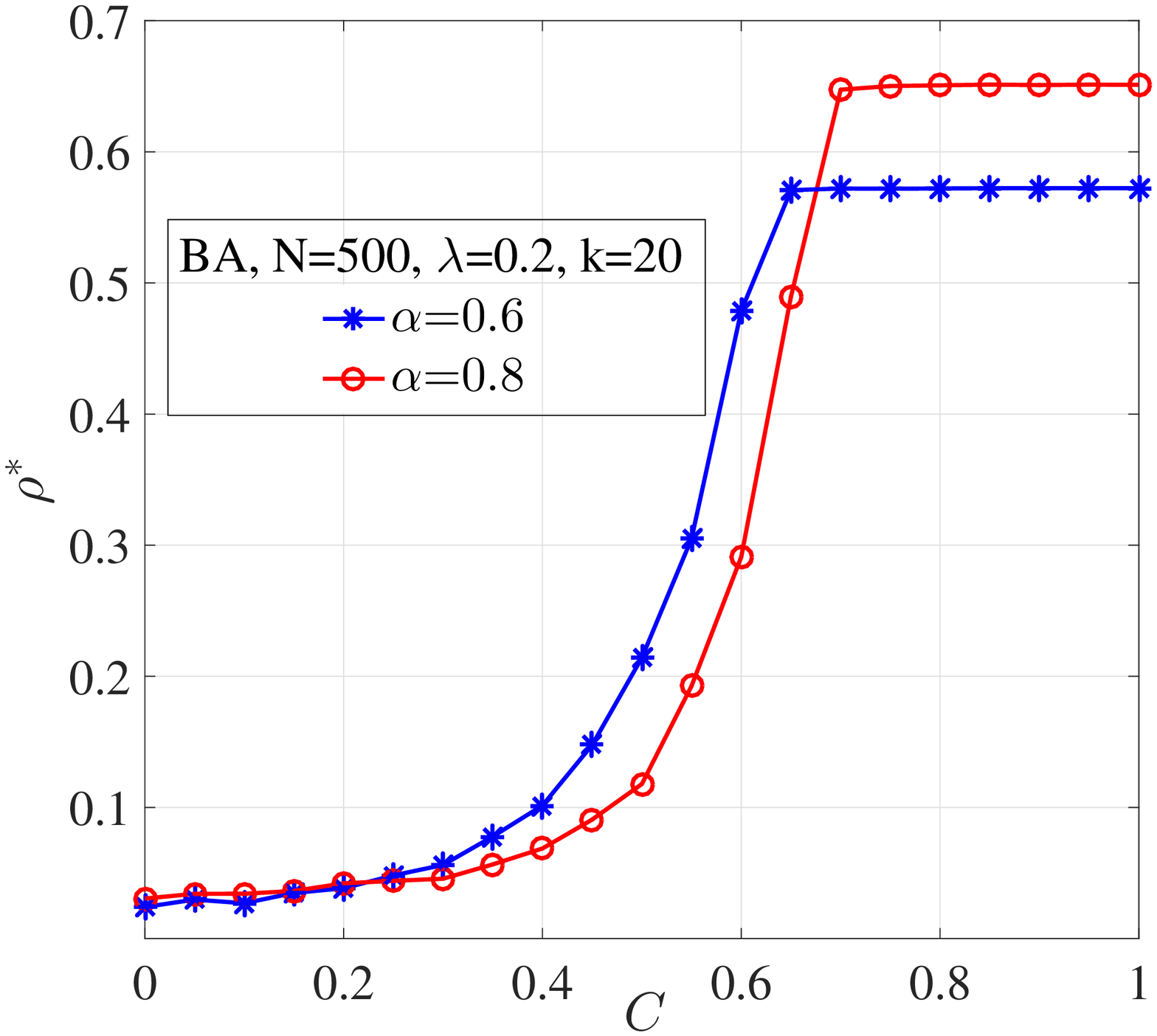}
\end{minipage}
}
\subfigure[~~~~~~]{
\begin{minipage}{0.3\linewidth}
\centering
\includegraphics[height=42mm,width=100mm]{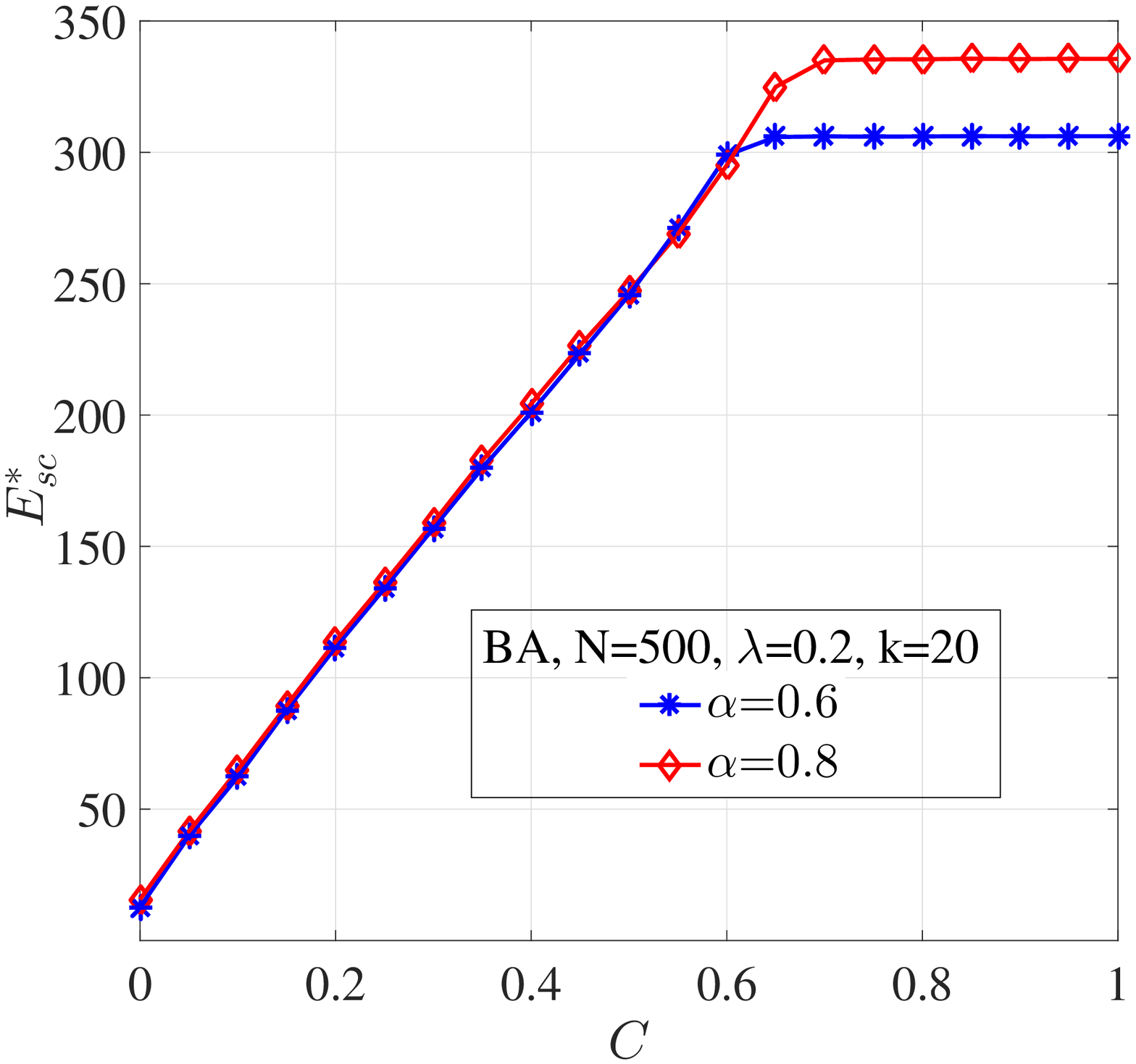}
\end{minipage}
}
\caption{For different strength coefficients, (a) vaccination equilibrium $x^*$, (b) final infection density $\rho^*$ and (c) social cost $E_{sc}^*$ as a function of vaccination cost $C$ in a two-layer BA scale-free network. $\mu=0.4$, $\beta=0.8$, $T=0.1$ and $\xi=0.5$.}
\end{figure*}

In order to study the vaccination performance in epidemic control, we consider that a proportion $x$ of individuals choose to be vaccinated in a two-layer ER network, and explore the impacts of vaccination on the infection density and social cost. Fig. 6 shows the social cost $E_{sc}$ and infection density $\rho^I$ as a function of vaccination density $x$, respectively. There exists a herd immunity threshold $x_c$ that results in the minimal social cost, which is consistent with the case of a single-layer network \cite{fu2011imitation}. When $x<x_c$, both infection density $\rho^I$ and social cost $E_{sc}$ decrease with the increase of vaccination density $x$. It is worth noting that the herd immunity threshold $x_c$ is not affected by the strength coefficient $\alpha$, but increases as the infection rate $\beta$ increases. Moreover, we find that the strength coefficient $\alpha$ has an influence on the social cost $E_{sc}$ and infection density $\rho^I$. As shown in Fig. 6 (a), since the vaccination density $x$ increases, the difference between the social costs $E_{sc}$ caused by different strength coefficients decreases. In summary, for a fixed vaccination density $x$, which is less than the herd immunity threshold $x_c$, more information may result in less social cost $E_{sc}$ and infection density $\rho^I$.

We now perform simulations on a two-layer ER network ($N=500$) to study the vaccination dynamics. The initial factions of vaccinated and infected individuals are set to 0.1 and 0.2, respectively. For a fixed communication cost $T=0.1$, Fig. 7 (a) shows that the vaccination equilibrium $x^*$ as a function of vaccination cost $C$ under the influence of information dissemination ($\alpha=0.6$). The solid line is calculated by Eq. (18). We find that there is a good agreement between the analytical solutions and numerical simulations, and vaccination equilibrium $x^*$ decreases with the increase of vaccination cost $C$. The introduced vaccination affects the epidemic propagation. Let $\rho^*$ denote the final infection density at the equilibrium. Figs. 7 (b) and 7 (c) illustrate that as the vaccination equilibrium $x^*$ increases, the final infection density $\rho^*$ decreases regardless of the network structure. In addition, the social cost at the equilibrium, denoted by $E_{sc}^*$, decreases with the decreases of final infection density $\rho^*$, as shown in the insets. Therefore, we conclude that the increase of vaccination equilibrium $x^*$ can effectively reduce the final infection density $\rho^*$ and social cost $E_{sc}^*$, and help to control the spread of epidemics.

How to improve the vaccination equilibrium motivates us to explore the impact factors of vaccination. For a fixed vaccination cost $C$, we compare the vaccination equilibrium $x^*$ with different strength coefficients to determine the effect of information dissemination intensity on the vaccination decision-making. Simulations are performed in a two-layer ER network and a two-layer BA scale-free network, respectively. As illustrated in Figs. 8 (a) and 9 (a), vaccination equilibrium $x^*=0$, when the vaccination cost is relatively large ($C>0.7$), the change of strength coefficient $\alpha$ has no effect on $x^*$. However, the vaccination equilibrium $x^*$ decreases with the decrease of strength coefficient $\alpha$, when $C\in (0.5,0.7)$ in Fig. 8 (a) and $C\in (0.2,0.7)$ in Fig. 9 (a). In other words, the increase of the intensity of information transmission leads to a decline of the vaccination equilibrium $x^*$, when the vaccination cost $C$ is within a certain range (of which upper bound depends on the network topology and lower bound is $\rho^I(0)+(1-\alpha)T$).

Figs. 8 (b) and 9 (b) show that with the increase of vaccination cost $C$, the final infection density $\rho^*$ increases. Similar to the procedure of $x^*$, we study how the strength coefficient $\alpha$ influences the final infection density $\rho^*$ with a given $C$. We find that when the vaccination cost $C$ is relatively small, the change of strength coefficient $\alpha$ has no effect on the final infection density $\rho^*$ as well. The final infection density $\rho^*$ under $\alpha=0.6$ is higher than the one under $\alpha=0.8$, when the vaccination cost $C$ is within a certain range. Associating the final infection density $\rho^*$ with the vaccination equilibrium $x^*$, we find that the final infection density $\rho^*$ increases when the information dissemination suppresses vaccination. When the vaccination cost $C$ is relatively large, vaccination equilibrium $x^*=0$, the opposite is true and the information dissemination can reduce the final infection density. In summary, although information transmission  can trigger some protective behaviors, e.g., wearing masks and washing hands, to reduce the effective infectivity at that season, the resulting reduction of vaccination density may lead to the increase of infection density at the next epidemic season. In terms of the social cost $E_{sc}^*$, only when the vaccination cost exceeds a certain value ($C>0.7$), the change of strength coefficient has an impact on the social cost (see Figs. 8 (c) and 9 (c)). As the strength coefficient increases, the social cost increases. In general, the effect of information dissemination on vaccination and epidemic control depends on the vaccination cost.

\begin{figure}[!htbp]
\centering
\includegraphics[height=100mm,width=85mm]{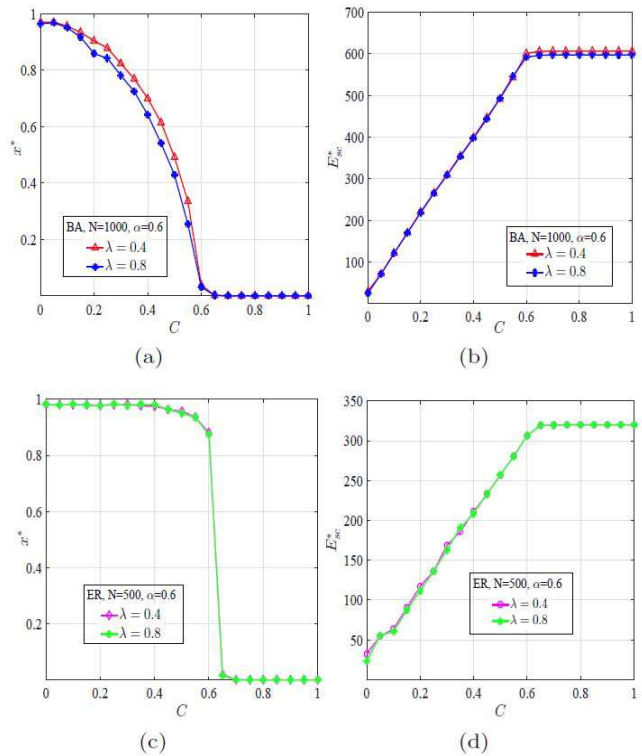}
\caption{Vaccination equilibrium $x^*$ and social cost $E_{sc}^*$ as a function of vaccination cost $C$ with different sensitivity coefficients $\lambda$ in (a)-(b) a two-layer BA scale-free network with $N=1000$ and (c)-(d) a two-layer ER network with $N=500$, respectively. $\mu=0.4$, $\beta=0.8$, $T=0.1$, $k=20$.  }
\end{figure}

We further explore the effect of sensitivity coefficient $\lambda$ on vaccination. For the sake of simplicity, we assume that each individual has the same sensitivity $\lambda$ to the information. We perform simulations on a two-layer BA scale-free network ($N=1000$) and a two-layer ER network ($N=500$), respectively. As illustrated in Figs. 10 (a) and 10 (b), the vaccination equilibrium $x^*$ decreases with the increase of sensitivity coefficient $\lambda$ when $C\in (0.2,0.65)$, indicating that information transmission cannot promote vaccination in a two-layer BA scale-free network. For a two-layer ER network, sensitivity coefficient has no effect on both vaccination equilibrium $x^*$ and social cost $E_{sc}^*$. Therefore, we conclude that the role of individual sensitivity to risk information in immune decision-making depends on the network topology.

\section{\label{sec_6}Conclusion}
Individuals behavioral responses to information dissemination determine its influence on epidemic control. Increasing the vaccination equilibrium can reduce the final infection density and social cost. Taking into account the complexity of decision-making of individuals in vaccination, we have presented an evolutionary vaccination game model by incorporating the information-epidemic propagation process into the vaccination dynamics, and explored the influence of information dissemination on vaccination. The impact of strength coefficient and individual sensitivity on the vaccination equilibrium reveals that more information transmission cannot promote vaccination. Information dissemination is negatively correlated with vaccination when vaccination cost $C<\rho^I(0)+(1-\alpha)T$. Although more information dissemination at one epidemic season can increase the epidemic threshold of that stage, it cannot improve the vaccination equilibrium in the whole multiple epidemic seasons. Therefore, during the whole process of the evolutionary vaccination game in multiple epidemic seasons, information transmission increases the final infection density. Since information dissemination is inevitable during the epidemic, the correlation between information dissemination and vaccination may provide a guidance for the authorities to implement information regulation for epidemic control. However, the situation of stochastic fluctuations that lead to the extinction of infection in finite networks \cite{hindes2018enhancement} has not been considered in this paper, which may be of interest in multiplex vaccination games in the future.

\begin{acknowledgments}
This work was partly supported by the National Natural Science Foundation of China (No. 71731004, No. 61603097), the National Natural Science Fund for Distinguished Young Scholar of China (No. 61425019), and Natural Science Foundation of Shanghai (No. 16ZR1446400).

\end{acknowledgments}

\end{document}